\documentclass[sigconf]{acmart}
\usepackage{tikz}
\usepackage{amsmath}
\usepackage{subfig}
\usepackage{amsmath}
\usepackage{cuted}
\usepackage{lipsum}
\usetikzlibrary{shapes.geometric, arrows.meta, positioning}

\newtheorem{remark}{\hskip -1em \bf \emph{\underline{Remark}}}
\newtheorem{property}{\hskip -1em \bf \emph{\underline{Property}}}

\newtheorem{corollary}{\hskip -1em \bf \emph{\underline{Corollary}}}
\newtheorem{lemma}{\hskip -1em \bf \emph{\underline{Lemma}}}

\def\qed{$\Box$}

\def\({\left(}
\def\){\right)}

\def\b0{{\mathbf{0}}}

\AtBeginDocument{%
  }

\setcopyright{acmlicensed}

\copyrightyear{2024} 
\acmYear{2024} 
\setcopyright{acmlicensed}\acmConference[ACM MobiCom '24]{The 30th Annual
International Conference on Mobile Computing and Networking}{November 18--22,
2024}{Washington D.C., DC, USA}
\acmBooktitle{The 30th Annual International Conference on Mobile Computing and
Networking (ACM MobiCom '24), November 18--22, 2024, Washington D.C., DC, USA}
\acmDOI{10.1145/3636534.3698231}
\acmISBN{979-8-4007-0489-5/24/11}


\begin{document}

\title{Near-Field Localization With Coprime Array}

\author{Hongqiang Cheng}
\email{chenghq2023@mail.sustech.edu.cn}
\affiliation{%
	\institution{Southern University of Science and Technology}
	\city{Shenzhen}
	\country{China}
}

\author{Changsheng You\textsuperscript{*}}
\email{youcs@sustech.edu.cn}
\affiliation{%
	\institution{Southern University of Science and Technology}
	\city{Shenzhen}
	\country{China}
}
 \thanks{\textsuperscript{*} Corresponding author}
\author{Cong Zhou}
\email{zhoucong@stu.hit.edu.cn}
\affiliation{%
	\institution{Harbin Institute of Technology}
	\city{Harbin}
	\country{China}
}

\begin{abstract}
    Large-aperture \emph{coprime arrays} (CAs) are expected to achieve higher sensing resolution than conventional dense arrays (DAs), yet with lower hardware and energy cost. However, existing CA far-field localization methods cannot be directly applied to near-field scenarios due to channel model mismatch. To address this issue, in this paper, we propose an efficient near-field localization method for CAs. Specifically, we first construct an effective covariance matrix, which allows to decouple the target angle-and-range estimation. Then, a customized two-phase multiple signal classification (MUSIC) algorithm for CAs is proposed, which first detects all possible targets' angles by using an angular-domain MUSIC algorithm, followed by the second phase to resolve the true targets' angles and ranges by devising a range-domain MUSIC algorithm. Finally, we show that the proposed method is able to locate more targets than the subarray-based method as well as achieve lower root mean square error (RMSE) than DAs. 
\end{abstract}


\maketitle

\section{Introduction}
Integrated sensing and communication (ISAC) has been recognized as a promising technology for future sixth-generation (6G) wireless networks, which enables both functions of sensing and communication by sharing the same spectrum and hardware resources~\cite{10663521,you2024generationadvancedtransceivertechnologies}. In particular, wireless/radar sensing has found a wide range of applications in practice, such as high-precision positioning and tracking, high-resolution imaging, and simultaneous localization and mapping. Moreover, with the technology trends evolving towards higher frequency band and larger transmitter array aperture~\cite{TutorialXLMIMO}, the Rayleigh distance, which specifies the boundary between the near-field and far-field, is greatly expanded, thus rendering environmental targets more likely to be located in the near-field region of the large-aperture array~\cite{10173734,9913211}. 

Compared with conventional far-field target localization, near-field localization with large-aperture arrays can achieve higher sensing resolution. However, new design challenges also need to be well addressed. For example, different from far-field localization based on planar-wavefronts, near-field localization needs to take into account the more accurate \emph{spherical}-wavefront model, which is characterized by both the target's angle and range. This thus makes the existing far-field oriented angle estimation methods inefficient when applied in near-field sensing systems.
To address this issue, various near-field localization methods have been recently proposed. For example, the authors in~\cite{2DMUSIC} developed a two-dimensional (2D) multiple signal classification (MUSIC) algorithm based on the the orthogonality between the signal subspace and noise subspace in near-field localization. To reduce the complexity of 2D exhaustive search, a low-complexity approach was proposed in~\cite{RDMUSIC}, where the authors designed an efficient reduced-dimension (RD) MUSIC algorithm transforming the 2D search into a one-dimensional (1D) local search.
In addition, the authors in~\cite{ramezani2024localizationmassivemimonetworks} presented a new method for target localization, which decoupled the angle and the range parameters by extracting anti-diagonal elements of the covariance matrix.
However, the above works mainly considered \emph{dense arrays} (DAs) with a large number of sensors/array elements, which practically incur high power and hardware cost, as well as demanding signal processing complexity.

To tackle this issue, \emph{sparse arrays} (SAs)~\cite{SparseMIMOforISAC} have been proposed as an alternative array configuration to achieve large array aperture with a small number of sensors only, hence greatly reducing the power consumption. In particular, apart from uniform SAs, structured SAs (such as coprime arrays (CAs)~\cite{SparseSensingWithCoPrime} and nested arrays (NAs)~\cite{NestedArrays}) have been also investigated to achieve superior localization performance. 
For example, it was shown in~\cite{VirtualArrayInterpolation} that for far-field localization, CAs endow higher sensing/localization degrees-of-freedom (DoFs) than DAs, which characterize the maximum number of sensing targets. This thus motivated growing research efforts recently to design efficient localization methods and beamforming designs tailored to CAs to achieve superior localization performance~\cite{zhou2024sparsearrayenablednearfield}. 
However, the existing far-field oriented CA localization methods cannot be directly applied to the near-field case, due to the spherical (instead of planar) channel model. One feasible solution is by decomposing the entire array into two symmetric SAs, hence eliminating the angular ambiguity by leveraging coprime properties.
Nevertheless, this approach fails to utilize the mutual information between the subarrays, thereby not fully exploiting the high sensing DoFs provided by CAs.

Motivated by the above, we study in this paper a new and efficient near-field localization method for large-aperture CAs. To this end, we first construct an effective covariance matrix for CAs, which allows to fully exploit all received signals and decouple the angle-and-range estimation. Then, a customized two-phase MUSIC algorithm for CAs is proposed, which first detects all possible targets' angles (including the true and cross angles) by using an angular domain MUSIC algorithm, followed by the second phase to resolve the true targets' angles and ranges by devising a range-domain MUSIC algorithm. Finally, we theoretically and numerically show that the proposed near-field target localization method is able to locate more targets than the conventional subarray-based method as well as achieve lower root mean square error (RMSE) than DAs.

\section{System model}
We consider a radar sensing system with a large-aperture (symmetric) CA, which is used for probing $K$ uncorrelated targets in the narrowband. In this section, the symmetric CA and near-field channel models are introduced.

\begin{figure}[!t]
	\centering
	\includegraphics[width=1\linewidth]{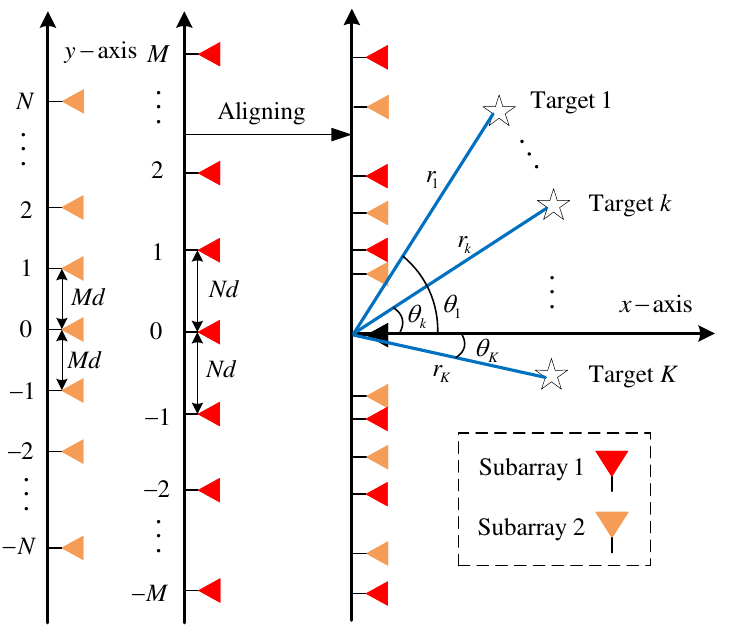}
	\caption{Illustration of the symmetric coprime array.}
	\label{sys}
 
\end{figure}

\textbf{\underline{Array model:}} As illustrated in Fig. \ref{sys}, the basic CA consists of two sparse ULAs with different inter-sensor spacing.  Specifically, subarray 1 is composed of $(2M-1)$ sensors with inter-sensor spacing of $Nd$, while subarray 2 is composed of $(2N-1)$ sensors with inter-sensor spacing of $Md$, where $d = \frac{\lambda}{4} $ is the minimum inter-sensor spacing with $\lambda$ denoting the carrier wavelength.
Moreover, ${M}$ and ${N}$ are coprime integers, and we assume that $M < N$ without loss of generality. 
Accordingly, the CA can be regarded as consisting of $U = 2V-1$ sensors in total, where $V = M + N -1$ denotes the number of sensors of the basic CA.
By assuming that the middle of the sensor is placed at the origin, the array sensors are positioned at  
\begin{equation}\begin{aligned}
		\label{eq:elements}
		\mathcal{S}&=\{Mnd~|~n=-N+1,-N+2,\cdots,N-1\}\\
		&\cup\{Nmd~|~m=-M+1,-M+2,\cdots,M-1\}.
\end{aligned}\end{equation}
Moreover, by rearranging the elements of $\mathcal{S}$ in an ascending order, we obtain an equivalent sensor location vector $\mathbf{s}=\left[s_1,s_2,\cdots,s_U \right]$.   

\textbf{\underline{Channel model:}} Let $(\theta_k,r_k),~k=1,2,\cdots,K$ denote the location of each target, where $\theta_k$ and $r_k$ denote the angle and range between the $k$-th target and the array center, respectively.
We assume that all targets are located in the Fresnel near-field region of the CA, where the BS-target range $r_k$ is larger than the Fresnel distance  $Z_F = \max\{d_{R},1.2D\}$ and smaller than the Rayleigh distance $Z_R = \frac{2D^2}{\lambda}$ with $D=2M(N-1)d$ denoting the array aperture. As $d_R$ is of several wavelengths~\cite{ouyang2024impactreactiveregionnearfield}, the Fresnel distance can be simplified as $Z_F=1.2D$. In this paper, we focus on the near-field sensing scenarios in high-frequency bands such as millimeter-wave (mmWave) and even terahertz (THz) for which the non line-of-sight (NLoS) channel paths have negligible power due to the severe path-loss and shadowing. 
Thus, we only consider the line-of-sight (LoS) path, whose channel steering vector can be defined as
\begin{equation}\label{eq1}\mathbf{b}(\theta_k,r_k)=\left[\mathrm{e}^{-\jmath\frac{2\pi}\lambda(d_{-V,k}-r_k)},\ldots,\mathrm{e}^{-\jmath\frac{2\pi}\lambda(d_{V,k}-r_k)}\right]^T,\end{equation}
with $d_{u,k},u\in\mathcal{U}\triangleq \{1,2,\cdots,U\}$ denoting the distance from the $u$-th sensor to the $k$-th target. Based on the Fresnel approximation~\cite{TutorialXLMIMO}, the phase delay of the $u$-th sensor can be approximated as 

\begin{equation}\begin{aligned}
		d_{u,k}-r_k &=\sqrt{r_k^2+s_u^2-2r_k s_u\sin\theta_k}-{r_k}  \\
		&\approx p_u(\theta_k)+q_u(\theta_k,r_k),
\end{aligned}\end{equation}
where $p_u(\theta_k)= -s_u 2\pi \sin\theta_k/\lambda$ and $q_u(\theta_k,r_k)=s^2_u\pi \cos^{2}\theta_k/\lambda r_k$.
Therefore, the channel steering vector in (\ref{eq1}) can be rewritten as
\begin{equation}
	\left[\mathbf{b}(\theta_k,r_k)\right]_u \approx e^{\jmath p_u(\theta_k)+\jmath q_u(\theta_k,r_k)}.
\end{equation}

\textbf{\underline{Signal model:}} For CA radar sensing, the received signal vector at the CA at time $t$ can be modeled as
\begin{equation}\begin{aligned}
		\mathbf{y}(t)=&\sum_{k=1}^K\mathbf{b}(\theta_k,r_k)x_k(t)+\mathbf{z}(t)\\
		=&\mathbf{B}(\mathbf{\theta},\mathbf{r})\mathbf{x}(t)+\mathbf{z}(t),
\end{aligned}\end{equation}
where $\mathbf{B}(\mathbf{\theta},\mathbf{r}) \triangleq \left[\mathbf{b}(\theta_{1},r_1),~\mathbf{b}(\theta_{2},r_2),~\cdots,~\mathbf{b}(\theta_{K},r_K)\right] \in \mathbb{C}^{U\times K}$ denotes the CA steering matrix and $\mathbf{x}(t) \triangleq \left[x_{1}(t),x_{2}(t),\cdots,x_{K}(t)\right]^{T}\in\mathbb{C}^{K}$ denotes the source vector. Moreover, $\mathbf{z}(t)\sim\mathcal{CN}(0,\sigma^{2}\mathbf{I})$ is the received circularly symmetric complex Gaussian (CSCG) noise vector with zero mean and variance $\sigma^{2}$.

\section{Proposed near-field localization algorithm}
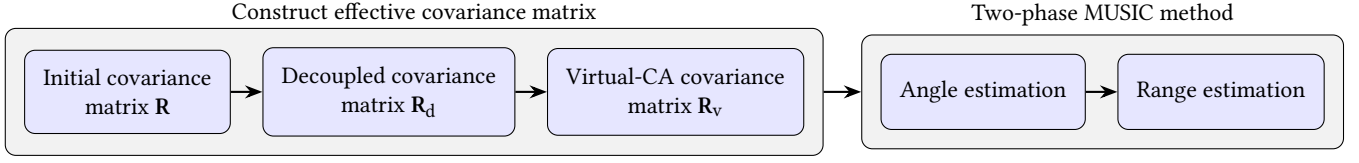
\begin{figure*}[htbp] 
	\centering
	
	\begin{tikzpicture}[
		node distance=0.5cm,
		box/.style={draw, fill=blue!10, minimum width=1.2cm, minimum height=1cm, align=center},
		group/.style={draw, fill=gray!10, rounded corners, inner sep=0.25cm},
		arr/.style={-Stealth, thick}
		]
		
		\node[group, label={above:Construct effective covariance matrix}] (group1) {
			\begin{tikzpicture}[node distance=0.42cm]
				\node[box] (box1) {Initial covariance\\ matrix $\mathbf{R}$};
				\node[box, right=of box1] (box2) {Decoupled covariance \\matrix $\mathbf{R}_\mathrm{d}$};
				\node[box, right=of box2] (box3) {Virtual-CA covariance \\matrix $\mathbf{R}_\mathrm{v}$};
				\draw[arr] (box1) -- (box2);
				\draw[arr] (box2) -- (box3);
			\end{tikzpicture}
		};
		
		\node[group, right=of group1, label={above:Two-phase MUSIC method}] (group2) {
			\begin{tikzpicture}[node distance=0.42cm]
				\node[box] (box4) {Angle estimation};
				\node[box, right=of box4] (box5) {Range estimation};
				\draw[arr] (box4) -- (box5);
			\end{tikzpicture}
		};
		
		\draw[arr] (group1.east) -- (group2.west);
		
	\end{tikzpicture}
	
	\caption{The framework of proposed near-field localization algorithm for CAs.}
	\label{fig:enter-label}
     \vspace{-12pt}
\end{figure*}
In this section, we propose a new and efficient near-field localization method for the CA, which is shown to achieve higher resolution than the conventional DA.

The main challenges of target localization for CAs are two-fold. 
Firstly, the angle estimation algorithms based on the virtual array in the far-field cannot be directly extended to the near-field case for the CA, due to the nonlinear phase delays across the array elements. 
Secondly, most existing near-field localization methods are designed for uniform linear arrays (ULAs). 
Although we can extend the ULA-based localization method to the CA by decomposing the CA into two ULAs, this approach suffers from degraded sensing DoFs due to the lack of inter-subarray mutual information (as will be numerically shown in Section 4).

To resolve these challenges, we first leverage the symmetry property of CAs to construct an effective covariance matrix. Then, a \emph{two-phase} near-field MUSIC method is proposed to successively estimate the angles and ranges of the targets. The main procedures of the algorithm are illustrated in Fig. 2.

\subsection{Construct Effective Covariance Matrix}
To facilitate the near-field localization, we first construct an effective covariance matrix tailored to the CA, which will be shown to enable effective decoupling of the near-field the angle and range parameters.

1) \textbf{\underline{Initial covariance matrix:}} 
Let $T$ denote the number of snapshots sent by the targets. 
First, the \emph{initial covariance matrix} $\mathbf{R}$ of received signals without noise taken into account at the CA is 
\begin{equation}\mathbf{R}\triangleq\mathbf{B}(\mathbf{\theta},\mathbf{r})\mathbf{B}^H(\mathbf{\theta},\mathbf{r}).
 \label{eq7}
\end{equation}
Since the covariance matrix is not available in practice, we consider its approximated version of the sample covariance matrix, which is given by 
\begin{equation}
	\hat{\mathbf{R}}=\frac{1}{T}\sum_{t=0}^{T-1}\mathbf{\hat{B}}_t\mathbf{\hat{B}}_t^{H},
 \label{eq8}
\end{equation}
To obtain the sampled covariance matrix in ~\eqref{eq8}, we use the least square (LS) estimation method, which can be expressed as
\begin{equation}\min_{\mathbf{\hat{B}}}~~\Vert\mathbf{y}(t)-\mathbf{\hat{B}}\mathbf{x}(t)\Vert_{\mathrm{F}}.\end{equation}
As such, the steering matrix $\mathbf{\hat{B}}_t$ at time $t$ can be obtained as
\begin{equation}\mathbf{\hat{B}}_t=\mathbf{y}(t)\mathbf{x}(t)\big(\mathbf{x}(t)^H\mathbf{x}(t)\big)^{-1}.\end{equation}

To facilitate the localization method design, we first analyze useful properties of the covariance matrix $\mathbf{R}\in\mathbb{C}^{U\times U}$. 
Specifically,  the $(i, j)$-th element of $\mathbf{R}$ in ~\eqref{eq7} is 
\begin{align}
        &[\mathbf{R}]_{i,j} =\sum_{k=1}^{K} \left[ \mathbf{b}(\theta_k,r_k)\right]_i [\mathbf{b}^H(\theta_k,r_k)]_j\nonumber  \\
        &=\sum_{k=1}^{K} \exp\left[p_i(\theta_k)-p_j(\theta_k)+q_i(\theta_k,r_k)-q_j(\theta_k,r_k)\right] \nonumber\\
        &=\sum_{k=1}^{K} \exp\left[-\jmath\frac{2\pi}{\lambda}((s_i-s_j)\sin\theta_k -\frac{(s_i+s_j)(s_i-s_j)\cos^2\theta_k}{2r_k})\right]\nonumber\\
        &=\sum_{k=1}^{K} \exp\left[P_{i,j}(\theta_k)+Q_{i,j}(\theta_k,r_k)\right],
\end{align}
where $P_{i,j}(\theta_k) \triangleq p_i(\theta_k)-p_j(\theta_k)$ and $Q_{i,j}(\theta_k,r_k) \triangleq q_i(\theta_k,r_k)-q_j(\theta_k,r_k)$ are defined as the \emph{first-order term} and the \emph{second-order term} of the covariance matrix element, respectively. Then we have the following results.

\begin{property}\rm
	(Symmetry of array position). For the symmetric CA, the sensors' position indices satisfy:
	\begin{equation}
		s_i = -s_{2V-i},~~\forall~i= 1,2,\cdots,U.
	\end{equation}
\end{property}
\begin{corollary}\rm
For anti-diagonal elements $[\mathbf{R}]_{i,2V-i}$, we have
\begin{align}
    [\mathbf{R}]_{i,2V-i}& = \sum_{k=1}^{K} \exp\left(-\jmath\frac{2\pi}{\lambda}2s_i\sin\theta_k\right)\nonumber\\
    & =\sum_{k=1}^{K} \exp\left[2P_{i,j}(\theta_k)\right].
\end{align}
\textbf{\emph{Corollary} 1} shows that the elements on the anti-diagonal of the initial covariance matrix contain the first-order terms only, which are dependent on the targets' angles solely. However, as there are only $U$ elements, using them alone for estimation cannot increase the DoFs of the CA.
\end{corollary}
\begin{property}\rm
 	(Symmetry of covariance matrix).
For the covariance matrix $\mathbf{R}$, consider the elements symmetric with respect to the anti-diagonal, i.e., $[\mathbf{R}]_{i,j}$ and $[\mathbf{R}]_{p,q}$, where $p=2V-j$ and $q=2V-i$. They have opposite second-order terms and identical first-order terms. Mathematically, we have 
\begin{equation}
    P_{p,q}(\theta_k)=P_{i,j}(\theta_k),\quad Q_{p,q}(\theta_k,r_k)=-Q_{i,j}(\theta_k,r_k).
\end{equation}

\begin{proof}
For the two elements $[\mathbf{R}]_{i,j}$ and $[\mathbf{R}]_{p,q}$, which are symmetric with respect to the anti-diagonal, by using \textbf{\emph{Property} 1} , we have
\begin{align}
    &[\mathbf{R}]_{p,q}=\sum_{k=1}^{K} \exp\left[P_{p,q}(\theta_k)+Q_{p,q}(\theta_k,r_k)\right]\nonumber \\
    &=\sum_{k=1}^{K} \exp\left[-\jmath\frac{2\pi}{\lambda}((s_p-s_q)\sin\theta_k 
-\frac{(s_p+s_q)(s_p-s_q)\cos^2\theta_k}{2r_k})\right]\nonumber\\
    &=\sum_{k=1}^{K} \exp\left[-\jmath\frac{2\pi}{\lambda}((s_i-s_j)\sin\theta_k +\frac{(s_i+s_j)(s_i-s_j)\cos^2\theta_k}{2r_k})\right]\nonumber\\
    &=\sum_{k=1}^{K} \exp\left[P_{i,j}(\theta_k)-Q_{i,j}(\theta_k,r_k)\right].
\end{align}

Therefore, it can be easily shown that for  $[\mathbf{R}]_{i,j}$ and $[\mathbf{R}]_{p,q}$, their first-order terms are equal, while their second-order terms are opposite, thus completing the proof.
\end{proof}
\end{property}

\begin{figure*}[ht]	
\begin{align}\label{eq2}   \relax[\mathbf{R}_\mathrm{d}]_{i,j}&=\sum_{k=1}^{K}\exp[P_{i,j}(\theta_k)+Q_{i,j}(\theta_k,r_k)]\times\sum_{k=1}^{K}\exp[P_{i,j}(\theta_k)-Q_{i,j}(\theta_k,r_k)]\nonumber\\
	&=\underbrace{\sum_{k=1}^{K}\exp[2P_{i,j}(\theta_k)]}_{\text{self-spectrum}}+ \underbrace{\sum_{w=1}^{K}\sum_{u=1,u\ne w}^{K}\exp[P_{i,j}(\theta_u)+P_{i,j}(\theta_w)+Q_{i,j}(\theta_u,r_u)-Q_{i,j}(\theta_w,r_w)]}_{\text{cross-spectrum}}\triangleq[\mathbf{S}]_{i,j}+[\mathbf{C}]_{i,j}.
\end{align}
    \rule{1\textwidth}{.4pt}
     \vspace{-15pt}
\end{figure*}

2) \textbf{\underline{Decoupled covariance matrix:}}
It is worth noting that for each covariance element in  $\mathbf{R}$, there are $K$ terms in the summation, each jointly determined by the target angle and range. This thus makes the classical 1D MUSIC algorithm inapplicable due to coupling of angle and range parameters. In order to fully exploit all elements in the initial covariance matrix for accurate target localization, rather than only using the anti-diagonal elements, we propose an efficient method to construct a \emph{decoupled} covariance matrix, which will facilitate the near-field target localization in the sequel. Specifically, by using \textbf{\emph{Property }2}, 
we define the \emph{decoupled covariance matrix} $\mathbf{R}_\mathrm{d}$ as 
\begin{equation}
	\mathbf{R}_\mathrm{d} = \mathbf{R}\odot \mathbf{R}_\mathrm{a},
\end{equation} 
where the new matrix $\mathbf{R}_\mathrm{a}$ is obtained by symmetry of the initial covariance matrix with respect to the anti-diagonal, i.e., $[\mathbf{R_\mathrm{a}}]_{i,j}=[\mathbf{R}]_{2V-j,2V-i}$, and $\odot$ denotes the Hadamard product.

\begin{lemma}\rm
The $(i, j)$-th element of decoupled covariance matrix $\mathbf{R}_\mathrm{d}$ is given by ~\eqref{eq2}, where $\mathbf{S}\in\mathbb{C}^{U\times U}$ is defined as the \emph{self-spectrum matrix} with each element given by
\begin{equation}
    [\mathbf{S}]_{i,j} = \sum_{k=1}^{K}\exp[2P_{i,j}(\theta_k)].
\end{equation}
In particular, $\mathbf{S}$ can be regarded as the far-field covariance matrix of CAs as follows
\begin{equation}
    \mathbf{S} = \mathbf{A}(\mathbf{\theta})\mathbf{A}^{H}(\mathbf{\theta}) = \sum_{k=1}^{K}\mathbf{a}(\theta_k)\mathbf{a}^{H}(\theta_k),
\end{equation}
where $\mathbf{A(\mathbf{\theta})}=[\mathbf{a}(\theta_{1}),\mathbf{a}(\theta_{2}),\cdots,\mathbf{a}(\theta_{K})]\in\mathbb{C}^{U\times K}$ with $\mathbf{a}(\theta_{k})$ defined as 
\begin{equation}
	\mathbf{a}(\theta_k)\!=\!\left[e^{-\jmath \frac{2\pi}{\lambda}2s_1 \sin\theta_k},e^{-\jmath \frac{2\pi}{\lambda}2s_2 \sin\theta_k},\cdots,e^{-\jmath \frac{2\pi}{\lambda}2s_U \sin\theta_k}\right]^{T}\!.
\end{equation}
In addition, $\mathbf{C}\in\mathbb{C}^{U\times U}$ is defined as the \emph{cross-spectrum matrix} with each element given by
\begin{align}
	[\mathbf{C}]_{i,j} &= \sum_{w=1}^{K}\sum_{u=1,u\ne w}^{K}\exp[P_{i,j}(\theta_u)+P_{i,j}(\theta_w) \nonumber\\
	&+Q_{i,j}(\theta_u,r_u)-Q_{i,j}(\theta_w,r_w)].
\end{align}
	
\end{lemma}

In the expression of $\mathbf{S}$, note that to avoid angle estimation ambiguity, the inter-sensor minimum spacing should satisfy $d \leq \lambda/4$.
Upon completing this phase, we obtain the self-spectrum, which contains only the first-order term. Meanwhile, this step also introduces the undesirable and extraneous cross-spectrum components. 

\begin{remark}\rm(Self-spectrum versus cross-spectrum).
Note that in \textbf{Lemma }1, $\mathbf{R}_\mathrm{d}$ is the sum of two parts, corresponding to the self spectrum and cross spectrum. This decoupled form will be used for firstly estimating the angles and then their ranges. In particular, the self-spectrum term $\mathbf{S}$ is determined solely by the angles of targets, thus it can form true spectral peaks corresponding to targets' angles when the classical MUSIC algorithm is applied. 
In contrast, the cross-spectrum term $\mathbf{C}$ is determined by the locations of every two targets, i.e., $(\theta_u, r_u)$ and $(\theta_w, r_w)$. Consequently, $\mathbf{C}$ corresponds to cross-spectral peaks that do not represent the true location of the target.
Therefore, it is important to distinguish true spectral peaks and cross spectral peaks for multi-target localization. 

\end{remark}

3) \textbf{\underline{Virtual covariance matrix:}}
Generally, the DoFs in target localization are constrained by the number of sensors. To leverage the enhanced DoFs provided by CAs and extend the far-field algorithm to the near-field scenario, we first vectorize the coupled covariance matrix $\mathbf{R}_\mathrm{d}$ to obtain an equivalent receiving signal of the virtual ULA as follows
\begin{align}
		\mathbf{r}_\mathrm{d}\triangleq\mathrm{vec}(\mathbf{R}_\mathrm{d})&=
		\mathrm{vec}(\mathbf{S})+\mathrm{vec}(\mathbf{C})\nonumber\\
		&=\mathbf{A}_\mathrm{v}(\theta)\mathbf{1}_\mathrm{K}+\mathbf{c},
\end{align}
where $\mathbf{A}_\mathrm{v}(\theta)=[\mathbf{a}^{*}(\theta_{1})\otimes\mathbf{a}(\theta_{1}),\mathbf{a}^{*}(\theta_{2})\otimes\mathbf{a}(\theta_{2}),\cdots,\mathbf{a}^{*}(\theta_{K})\otimes\mathbf{a}(\theta_{K})] \\ \in\mathbb{C}^{U^{2}\times K}$, $\mathbf{1}_\mathrm{K}=[1,1,\cdots,1]^T\in\mathbb{C}^{K}$, $\mathbf{c}=\text{vec}(\mathbf{C})$, and $\otimes$ denotes the Kronecker product. 
The vector $\mathbf{r}_\mathrm{d}$ can be regarded as an equivalent received signal of the virtual array, corresponding to the steering matrix $\mathbf{A}_\mathrm{v}$, with the virtual sensors located at \begin{equation}
    \mathcal{S}_\mathcal{V}=\{(s_m-s_n)|m,n=0,1,\cdots,U \}.
\end{equation}
However, note that the covariance matrix obtained directly from the equivalent virtual signal is a rank-one matrix, i.e. $\text{rank}(\mathbf{r}_\mathrm{d}\mathbf{r}_\mathrm{d}^H)=1$. Therefore, the targets' angles can hardly be identified from $\mathbf{r}_\mathrm{d}$ when there are multiple targets. To address this issue, we apply the spatial smoothing technique for multiple targets sensing~\cite{5739227}. Specifically, we first construct a matrix 
$\mathbf{\tilde{A}}=[\tilde{\mathbf{a}}(\theta_{1}),\tilde{\mathbf{a}}(\theta_{2}),\cdots,\tilde{\mathbf{a}}(\theta_{K})]\in\mathbb{C}^{(2MN+1)\times K}$ by removing repeated rows in the matrix $\mathbf{A}_\mathrm{v}$ and sorting the consecutive remaining rows, so that the rows in $\mathbf{\tilde{A}}$ are identical to a consecutive virtual ULA. For symmetric CAs, there exist at least $(2MN+1)$ consecutive and unique virtual array elements~\cite{5739227}. 
As such, the vector $\mathbf{r}_\mathrm{d}$ can be transformed as \begin{equation}\mathbf{\tilde{r}}=\mathbf{\tilde{A}}\mathbf{1}_\mathrm{K}+\mathbf {\tilde{c}}.\end{equation}
Then, the consecutive virtual ULA can be divided into $(MN+1)$ overlapping subarrays with $(MN+1)$ virtual sensors for each subarray. The equivalent received signal vector of the $i$-th virtual subarray can be denoted as
\begin{equation}{\mathbf{\tilde{r}}}_{i}=\mathbf{\tilde{A}}_{i}\mathbf{1}_\mathrm{K}+\mathbf{\tilde{c}}_{i}.\end{equation}
The spatially smoothed covariance matrix can then be obtained by averaging the covariance matrix of each subarray over the $(MN+1)$ subarrays as
\begin{equation}\mathbf{R}_\mathrm{v}=\frac{1}{MN+1}\sum_{i=1}^{MN+1}\mathbf{\tilde{R}}_i=\frac{1}{MN+1}\sum_{i=1}^{MN+1}\mathbf{\tilde{r}}_{i}\mathbf{\tilde{r}}_{i}^H,\end{equation}
where $\mathbf{R}_\mathrm{v}$ is the so-called \emph{virtual-CA covariance matrix}. Note that virtual-CA covariance matrix $\mathbf{R}_\mathrm{v}$ is now full-rank which allows to identify up to $(MN+1)$ peaks in the spectrum, which will be applied in the following steps.

\subsection{Two-Phase MUSIC Method}
In this section, based on constructed effective covariance matrix, we propose an efficient two-phase MUSIC localization method to successively estimate the targets' angles and ranges with CAs. 

 \textbf{\underline{Phase 1: Angle estimation:}}
First, for the obtained effective covariance matrix $\mathbf{R}_\mathrm{v}$, one can observe that it contains decoupled angular information. Therefore, in the first stage, we perform a 1D spectral peak search in the angular domain for this effective matrix, which essentially includes both the true angles and undesirable cross angles. To this end, we first apply the eigenvalue decomposition to $\mathbf{R}_\mathrm{v}$
\begin{equation}\mathbf{R}_\mathrm{v}=\tilde{\mathbf{U}}_\mathrm{s}\tilde{\boldsymbol{\Sigma}}_\mathrm{s}\tilde{\mathbf{U}}_\mathrm{s}^H+\tilde{\mathbf{U}}_\mathrm{n}\tilde{\boldsymbol{\Sigma}}_\mathrm{n}\tilde{\mathbf{U}}_\mathrm{n}^H,\end{equation}
where $\tilde{\mathbf{U}}_\mathrm{s}$ and $\tilde{\mathbf{U}}_\mathrm{n}$ denote the signal and noise subspaces of the virtual covariance matrix, respectively. The possible angles can be estimated by searching the peaks from the spectrum
\begin{equation}
\hat{\mathbf{\theta}}=~^L\underset{\theta}{\operatorname*{argmax}}\frac{1}{\tilde{\mathbf{a}}^H(\theta)\tilde{\mathbf{U}}_\mathrm{n}\tilde{\mathbf{U}}_\mathrm{n}^H\tilde{\mathbf{a}}(\theta)}.
\end{equation}
Let $L$ denote the number of detected angles
$\left[\hat\theta_1,\hat\theta_2,\cdots,\hat\theta_L\right]$ from the spectral peaks, which includes both the true targets' angles and the cross targets' angles.

\textbf{\underline{Phase 2: Range estimation:}}
In the second phase, we propose an efficient method to identify the cross targets' angles and the true targets' angles, while obtaining accurate range estimation corresponding to the true targets' angles. The key idea is to utilize the fact that cross angles do not exhibit obvious spectrum peaks in the range domain, hence allowing for the detection of true targets. This phase is grounded in the theory that the near-field steering matrix $\mathbf{B}(\mathbf{\theta},\mathbf{r})$ and noise subspace matrix ${\mathbf{U}}_\mathrm{n}$ are orthogonal in the polar domain, i.e., $\mathbf{B}^H(\mathbf{\theta},\mathbf{r}){\mathbf{U}}_\mathrm{n}=\mathbf{0}$. 
With the detected angles, we re-express the receiving signal as
\begin{equation}\mathbf{Y}=\mathbf{B}(\hat{\mathbf{\theta}},\mathbf{r})\mathbf{x}+\mathbf{n}.\end{equation}
We distinguish cross angle and estimate the range parameter based on the initial covariance matrix by using the 1D MUSIC algorithm in the range domain. The initial covariance matrix of $\hat{\mathbf{R}}$ is given by $\hat{\mathbf{R}}=\mathbf{U}_\mathrm{s}\mathbf{\Sigma}_\mathrm{s}\mathbf{U}_\mathrm{s}^{H}+\mathbf{U}_\mathrm{n}\mathbf{\Sigma}_\mathrm{n}\mathbf{U}_\mathrm{n}^{H}$, where  ${\mathbf{U}}_\mathrm{s}$ and ${\mathbf{U}}_\mathrm{n}$ denote the signal and noise subspaces of the initial covariance matrix, respectively. Then, the range of detected  angles $\hat{\theta}_k$ can be obtained by finding the peaks from the spectrum below
\begin{equation}
	\hat{r}=\underset{r}{\operatorname*{argmax}}\frac1{\mathbf{b}^H(\hat{\theta}_k,r)\mathbf{U}_\mathrm{n}\mathbf{U}_\mathrm{n}^H\mathbf{b}(\hat{\theta}_k,r)}.
\end{equation}
For a detected angle $\hat{\theta}_k$, there are two possible cases in the process of range domain spectrum peak searching.
\begin{itemize}
    \item When the range spectrum exhibits no significant peaks across the entire range domain, the corresponding angle is identified as a cross angle.
    \item When $\hat{r}\in[1.2D,2D^{2}/\lambda]$, the corresponding angle is identified as a true angle.  
\end{itemize}
 It should be noted that the performance of range estimation is affected by the accuracy of angle estimation.

\begin{remark}\rm (Maximum number of estimable targets).
From the analysis of coupled covariance matrix components above, the maximum number of targets that the proposed method can estimate is mainly dependent on two factors.
Firstly, for $K$ near-field targets, we obtain $K$ true angles, which provide the true angle information of the targets. 
In addition, the cross-spectra also occupy the DoFs of the virtual array (for symmetric CAs, $\text{DoFs}\geq(MN+1)$~\cite{5739227}), thereby reducing the number of estimable targets. Given that cross-spectra occur for each pair of targets, the total number of cross-spectra is 
$K(K+1)/2$. As such, the maximum number of targets $K_{\mathrm{v}}$ that can be estimated satisfies the following
\begin{equation}
    K_{\mathrm{v}}+\frac{K_{\mathrm{v}}(K_{\mathrm{v}}-1)}{2} = MN+1.
\end{equation}
Further, $K_{\mathrm{v}} = \lfloor \sqrt{2MN+\frac{9}{4}}-\frac{1}{2} \rfloor\approx \sqrt{2MN}$, where $\lfloor~~ \rfloor$ denotes the rounding down function. On the other hand, it was shown in~\cite{VirtualArrayInterpolation} the maximum number of estimable targets using the subarray-based method of the CA is $K_{\mathrm{p}} = \text{min} \{M,N\}$. Consequently, our method significantly enhances the number of estimable targets, achieving nearly a $1.5$-fold increase compared to the subarray-based method. 
Specifically, as the difference between M and N increases, the improvement of our proposed method over the subarray-based method becomes more significant.

\end{remark}

\section{Numerical result}
\begin{figure}[t]
	\centering
	\subfloat[Angle spectrum]{
		\includegraphics[scale=0.29]{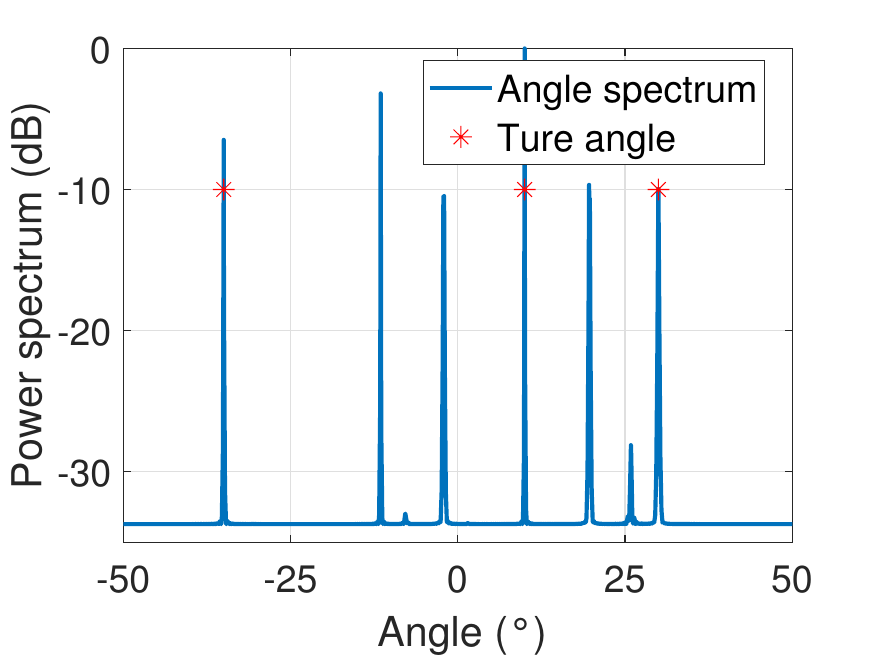}}
	\subfloat[Range spectrum]{
		\includegraphics[scale=0.29]{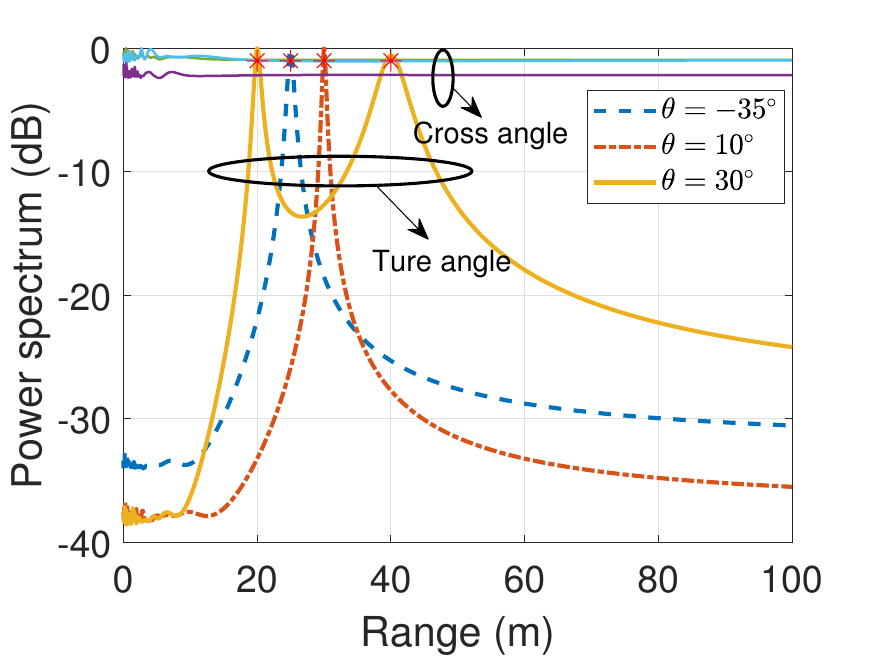}}
	\caption{Detectable angle and range spectrum.}
	\label{fig_2}
\end{figure}

Numerical results are provided in this section to validate the effectiveness of proposed CA near-field localization scheme. The system setup is as follows without otherwise specified. We consider a large-aperture CA with $M = 9$ and $N = 11$, which operates at the frequency band of $f = 30$ GHz. The number of snapshots is set as $T=100$. There are $K = 4$ targets with the angle-range pairs given by (-35°, 25 m), (10°, 30 m), (30°, 20 m), (30°, 40 m). This setup represents a challenging radar sensing case where two targets are located at the same angle but different ranges. For performance comparison, we consider the following benchmark schemes: 
1) DA based near-field sensing~\cite{ramezani2024localizationmassivemimonetworks}, where we assume the same number of sensors at the CA and DA; 2) far-field oriented CA sensing method which essentially utilizes the virtual array method in~\cite{5739227}; and 3) subarray-based near-field CA sensing method, where two uniform SAs separately are utilized for localization. Moreover, we consider the RMSE for the angle and range estimation as performance matrices, which are defined as 
\begin{small}\begin{equation}
	\theta_{\mathrm{RMSE}}=\sqrt{\frac{1}{QK}\sum_{q=1}^{Q}\sum_{k=1}^{K}\left(\hat{\theta}_k^{(q)}-\theta_{k}\right)},
	r_{\mathrm{RMSE}}=\sqrt{\frac{1}{QK}\sum_{q=1}^{Q}\sum_{k=1}^{K}\left(\hat{r}_k^{(q)}-r_{k}\right)},
\end{equation}\end{small}
where $Q=100$ stands for number of Monte Carlo simulations.
{\small{$\hat{\theta}_k^{(q)}$}} and {\small{$\hat{r}_k^{(q)}$}} denote the estimated angle and range for target $k$ in the $q$-th Monte Carlo simulation, respectively. 

Fig. \ref{fig_2}(a) shows the results of spectral peak search in the angular domain by using the CA, where the true targets locations are marked by red stars. It is observed that
by using the the first-phase MUSIC algorithm, the true angles of the targets can be well detected, while some undesired cross angles also appear. Next, we show in Fig. \ref{fig_2}(b) the results of spectral peak search in the range domain along all detected angles. It is observed that the second-phase MUSIC algorithm can well distinguish the true and cross targets' angles. In particular, there exist high  spectral peaks at the true angles only, while at the cross angles, no significant spectral peaks appear within the considered near-field region. Furthermore, one can observe that even when two targets are positioned at the same angle, they can still be distinguished in the range domain.

\begin{figure}[t]
	\centering
	\includegraphics[width=1\linewidth]{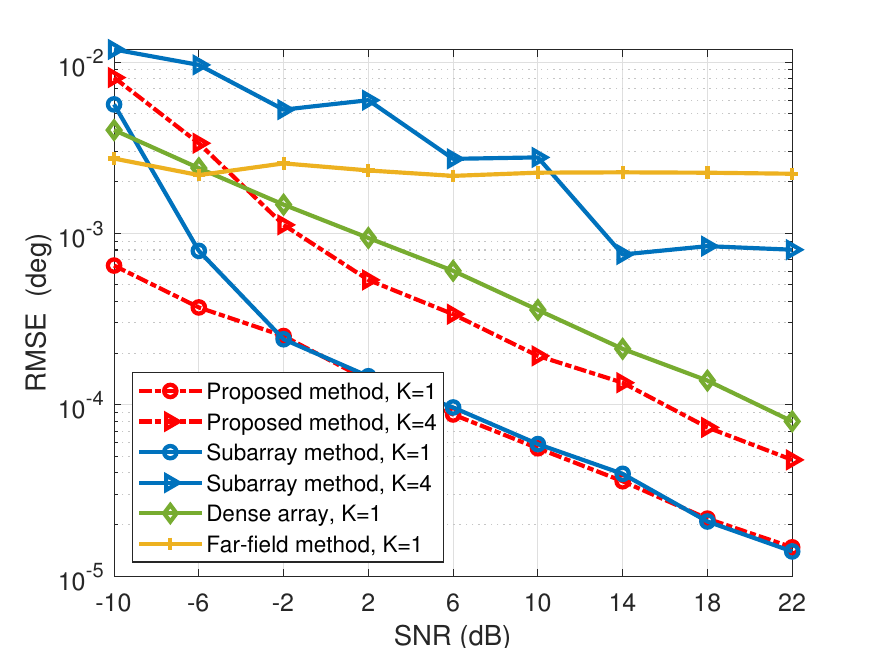}
	\caption{Angle RMSE versus SNR.}
	\label{fig_3}
\end{figure}

In Fig. \ref{fig_3}, we plot the curves of the angle RMSE versus the transmitted signal-to-noise (SNR) by different schemes. It is observed that the considered CA achieves better performance than the DA in the angle estimation. This is expected, as with the same number of sensors, CAs form a larger aperture than DAs, improving sensing resolution and accuracy. As SNR increases, both CAs and DAs attain smaller RMSEs. On the other hand, the far-field based method suffers a large RMSE even at high SNR. This indicates that it is improper to directly apply the virtual array method to near-field target localization. Moreover, as the number of targets increases, the proposed method is shown to be more effective compared to the subarray-based method. This is because for the latter, each uniform SAs exhibits periodic spectra during estimation. As the number of spectral peaks grows significantly, further complicating the process of obtaining accurate common spectral peaks.

\begin{figure}[tbp]
	\centering
	\includegraphics[width=1\linewidth]{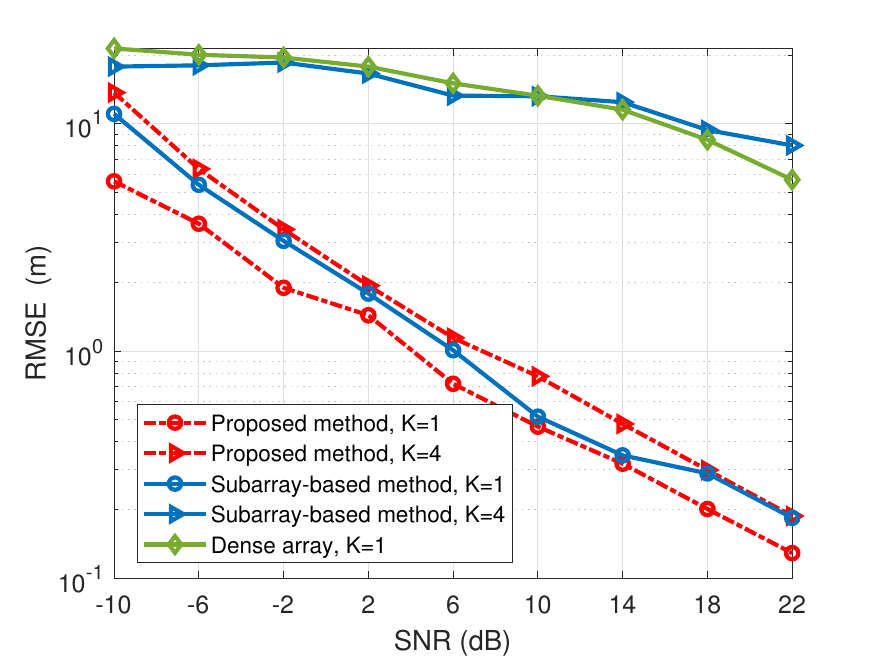}
	\caption{Range RMSE versus SNR.}
	\label{fig_4}
\end{figure}

Last, Fig. \ref{fig_4} shows the range RMSE versus SNR for different schemes. It is observed that as SNR increases, the range RMSE by the CA scheme decreases more significantly than DAs. This is due to the fact that for arrays with the same number of sensors, CAs are more likely to form a near-field region. The range information contained within this region is more prominent, leading to significantly higher precision for CAs compared to DAs. 
Last, one can observe that as the number of targets increases, the subarray-based method fails to achieve accurate range estimation, whereas the proposed method remains effective in providing accurate target localization.

\section{Conclusions}
In this paper, we proposed a near-field localization method by utilizing large-aperture CAs. 
Firstly, we analyzed the properties of the received signal covariance matrix in the near-field and constructed an effective covariance matrix. Then, we implemented a two-phase MUSIC method to accurately estimate the target location.
Finally, numerical results were presented to demonstrate the effectiveness of proposed method.

\begin{acks}
This work was supported in part by the National Key R\&D Program Youth Scientist Project under Grant 2023YFB2905100,  in part by the National Natural Science Foundation of China under Grant 62201242, 62331023, in part by Natural Science Foundation of Guangdong Province under Grant 2024A1515010097, in part by Shenzhen Science and Technology Program under Grant 20231115131633001.
\end{acks}

\bibliographystyle{ACM-Reference-Format}
\bibliography{sample}

\appendix

\end{document}